\documentclass{article}
\usepackage{amsmath}
\usepackage{graphicx}
\usepackage{amssymb}
\usepackage{amsfonts}
\usepackage{latexsym,epsfig}

\input{tcilatex}

\begin{document}

\title{Quantum Particle Production \\
at Sudden Singularities }
\author{John D. Barrow$^{1}$, Antonio B. Batista$^{2}$, \and J\'{u}lio C.
Fabris$^{2,3}$ and St\'{e}phane Houndjo$^{2}$ \\
$^{1}$DAMTP, Centre for Mathematical Sciences,\\
Cambridge University, Wilberforce Road,\\
Cambridge CB3 0WA, UK\\
$^{2}$Departamento de F\'{\i}sica, \\
Universidade Federal do Esp\'{\i}rito Santo, \\
CEP 29060-900 Vit\'{o}ria, Esp\'{\i}rito Santo, Brasil\\
$^{3}$ $Gr\epsilon CO$, Institut d'Astrophysique de Paris - IAP, \\
98bis, Boulevard Arago, 75014 Paris, France}
\date{}
\maketitle

\begin{abstract}
We investigate the effects of quantum particle production on a classical
sudden singularity occurring at finite time in a Friedmann universe. We use
an exact solution to describe an initially radiation-dominated universe that
evolves into a sudden singularity at finite time. We calculate the density
of created particles exactly and find that it is generally much smaller than
the classical background density and pressure which produce the sudden
singularity. We conclude that, in the example studied, the quantum particle
production does not lead to the avoidance or modification to the sudden
future singularity. We argue that the effects of small residual anisotropies
in the expansion will not change these results and show how they can be
related to studies of classical particle production using a bulk viscosity.
We conclude that we do not expect to see significant observable effects from
local sudden singularities on our past light cone.
\end{abstract}

\vspace{0.5cm} \leftline{PACS: 98.80.-k, 04.62.+v}

\section{\protect\bigskip Introduction}

In ref. \cite{jdb1} it was shown that classical general relativistic
Friedmann universes allow finite-time singularities to occur in which the
scale factor, $a(t)$, its time derivative, $\dot{a}$, and the density, $\rho
,$ remain finite whilst a singularity occurs in the fluid pressure, $%
p\rightarrow +\infty ,$ and the expansion acceleration, $\ddot{a}\rightarrow
\infty $. Remarkably, the strong energy condition continues to hold, $\rho
+3p>0.$ Analogous solutions are possible in which the singularity can occur
only in arbitrarily high derivatives of $a(t),$\cite{jdb2}. This behaviour
occurs independently of the 3-curvature of the universe and can prevent
closed Friedmann universes that obey the strong energy condition from
recollapsing \cite{jdb3}. Subsequently, a number of studies have been
carried out which generalise these results to different cosmologies and
theories of gravity \cite{jdb4}, classify the other types of future
singularity that can arise during the expansion of the universe \cite{odin},
and explore some observational constraints on their possible future
occurrence in our Universe \cite{dab}.

In this paper we extend these studies to consider some quantum implications
of a sudden singularity. Specifically, we want to know if quantum particle
production can dominate over the classical background density on approach to
a sudden singularity and stop it occurring or modify its properties, as can
be the case for the Big Rip future singularities \cite{qrip, no}. The
results of such a study are also of interest for observational probes of
finite-time singularities. In an inhomogeneous universe it would be possible
for us in principle to observe a sudden singularity on our past light cone.
What might we see? If quantum effects produce profuse particle production
then there might be observable effects from local sudden singularities. In
this respect, sudden singularities are also of interest with regard to
proposed measures of gravitational entropy associated with invariants of the
Weyl curvature \cite{weyl}. The Weyl invariant will not diverge on approach
to a sudden singularity (and there is no geodesic incompleteness \cite{geo})
and so it may represent a part of a soft future boundary of the universe
with low gravitational entropy -- which could be as close as 8.7 Myr to the
future \cite{dab}. The effects of loop quantum gravity have been studied in
cosmologies exhibiting classical sudden singularities and they can remove
the sudden singularity under certain particular conditions \cite{loop}, and
there is a close relationship between sudden singularities and the behaviour
of Friedmann universes containing bulk viscous stresses \cite{visc}.

In order to provide some insights into these issues we will construct a
simple exact classical Friedmann cosmological model with a future sudden
singularity that follows an era of radiation domination in which there is no
quantum production of massless particles. We will calculate the quantum
production of massless particles on approach to the sudden singularity where
there is a departure from conformal invariance. For simplicity, we will
ignore any period of late-time acceleration in the universe although this
more realistic detail can easily be incorporated. We will then discuss
various elaborations of this model and show why we do not expect small
deviations from isotropy and homogeneity to alter our conclusions.

\section{A Suddenly Singular Cosmology}

We will employ the simplest example of a spatially-flat isotropic Friedmann
universe with a sudden singularity, presented in \cite{jdb1}. The
cosmological equations for scale factor are: 
\begin{eqnarray}
3\biggr(\frac{\dot{a}}{a}\biggl)^{2} &=&8\pi G\rho ,  \label{1} \\
\frac{\ddot{a}}{a} &=&-\frac{4\pi G}{3}(\rho +3p),  \label{2} \\
\dot{\rho}+3\frac{\dot{a}}{a}(\rho +p) &=&0.  \label{3}
\end{eqnarray}%
We take the following solution for the scale factor and its first and second
derivatives: 
\begin{eqnarray}
a &=&\biggr(\frac{t}{t_{s0}}\biggl)^{q}(a_{s0}-1)+1-\biggr(1-\frac{t}{t_{s0}}%
\biggl)^{n},  \label{a} \\
\dot{a} &=&\frac{q}{t_{s0}}\biggr(\frac{t}{t_{s0}}\biggl)^{q-1}(a_{s0}-1)+%
\frac{n}{t_{s0}}\biggr(1-\frac{t}{t_{s0}}\biggl)^{n-1},  \label{b} \\
\ddot{a} &=&\frac{q(q-1)}{t_{s0}^{2}}\biggr(\frac{t}{t_{s0}}\biggl)%
^{q-2}(a_{s0}-1)-\frac{n(n-1)}{t_{s0}^{2}}\biggr(1-\frac{t}{t_{s0}}\biggl)%
^{n-2}.
\end{eqnarray}%
Imposing $0<q\leq 1$ and $1<n<2$, we see that $a\rightarrow a_{s0}$, $\dot{a}%
\rightarrow \frac{q(a_{s0}-1)}{t_{s0}}$ and $\ddot{a}\rightarrow \infty $,
as $t\rightarrow t_{s0}$. Hence, the scale factor and its first derivative
remain finite, implying that the density remains also finite, while the
second derivative of the scale factor, and consequently the pressure,
diverges. We assume the condition $0<q<1$ in order to have a decelerating
universe as $t\rightarrow 0$, but this condition could be relaxed without
affecting our results.

There is no exact solution for the Klein-Gordon equation for the form (\ref%
{a}) for the scale factor. Therefore, we consider some simplifications. We
divide the entire evolution of the universe into two phases: one which
characterizes the primordial phase, $t\rightarrow 0,$ and other which
characterizes the "singular" phase, $t\rightarrow t_{s0}$. For the
primordial phase, we will use the radiation-dominated phase of the standard
model (i.e. $q=1/2$), for which it is possible to solve the Klein-Gordon
equation so that the solution naturally contains the structure of the vacuum
state of quantum fields.

Let us consider the asymptotic behaviours of the solution (\ref{a}):

\begin{itemize}
\item Primordial radiation phase ($t\rightarrow 0$): 
\begin{eqnarray}
a &=&\frac{a_{s0}-1}{t_{s0}^{1/2}}t^{1/2}, \\
\dot{a} &=&\frac{(a_{s0}-1)}{2\,t_{s0}^{1/2}}t^{-1/2}.
\end{eqnarray}

\item Singular phase ($t\rightarrow t_{s0}$): 
\begin{eqnarray}
a &=&a_{s0}, \\
\dot{a} &=&\frac{(a_{s0}-1)}{2t_{s0}}.
\end{eqnarray}
\end{itemize}

We introduce the conformal time, $\eta $, defined by $ad\eta =dt$ and
re-express the scale factor during the radiative phase as $a=r\,t^{\frac{1}{2%
}}$, where the constant $r=\frac{a_{s0}-1}{\sqrt{t_{s0}}}$, so $\eta =\frac{2%
}{r}t^{\frac{1}{2}}.$This leads finally to the following expressions:

\begin{itemize}
\item Primordial radiation phase ($\eta \rightarrow 0$): 
\begin{eqnarray}
a &=&a_{0}\eta , \\
a^{\prime } &=&a_{0}.
\end{eqnarray}

\item Singular phase ($\eta \rightarrow \eta _{s0}$): 
\begin{eqnarray}
a &=&a_{s0}, \\
a^{\prime } &=&\frac{a_{0}a_{s0}}{2\sqrt{t_{s0}}}.
\end{eqnarray}
\end{itemize}

In the above expressions, $a_{0}=\frac{r^{2}}{2}$ and $^{\prime }=d/d\eta $.
The transition to the singular phase now occurs at the conformal time $\eta
=\eta _{0}$ and we have 
\begin{equation}
a_{p}(\eta _{0})=a_{s}(\eta _{0}),\quad a_{p}^{\prime
}(\eta_{0})=a_{s}^{\prime }(\eta _{0}),
\end{equation}
where the subscripts $p$ and $s$ denote \textit{primordial} and \textit{%
singular phases}, respectively. This implies that 
\begin{equation}
\eta _{0}=\frac{a_{s0}}{a_{0}}=\frac{2a_{s0}t_{s0}}{(a_{s0}-1)^{2}}\quad .
\end{equation}%
The matching conditions imply 
\begin{equation}
a_{0}=\frac{a_{s0}}{\eta _{0}},\quad a_{s0}=2\sqrt{t_{s0}},
\end{equation}%
and the solutions for the two phases are:

\begin{enumerate}
\item Primordial radiation phase: 
\begin{equation}
a=a_{s0}\frac{\eta }{\eta _{0}},\quad a^{\prime }=a_{0}=\frac{a_{s0}}{\eta
_{0}};
\end{equation}

\item Singular phase: 
\begin{equation}
a=a_{s0},\quad a^{\prime }=a_{0}=\frac{a_{s0}}{\eta _{0}}.
\end{equation}
\end{enumerate}

In order to construct the simplified model of approach to a sudden
singularity, we will consider $\eta _{0}$, $\eta _{s}$ and $a_{s0}$ to be
independent free parameters.

\section{The Klein-Gordon equation}

In a spatially-flat Friedmann universe, the Klein-Gordon equation for a
massless field, $\phi $, is 
\begin{equation}  \label{kge}
\Box \phi \equiv \phi ^{\prime \prime }+2\frac{a^{\prime }}{a}\phi ^{\prime
}+k^{2}c^{2}\phi =0,
\end{equation}%
where we have re-inserted the light velocity, $c,$ and if $k$ is the wave
number of the Fourier decomposition: 
\begin{equation}
\phi (\eta ,\vec{x})=\frac{1}{(2\pi )^{3/2}}\int \phi _{k}(\eta )e^{-i\vec{k}%
\cdot \vec{x}}d^{3}k.
\end{equation}%
For simplicity, we have omitted the Fourier index in the function $\phi $.
Now, we scale by $H_{0}^{2}$, where $H_{0}$ is the Hubble parameter at the
moment of the transition, $\eta _{0}$. Moreover, we define $\bar{k}=k\,l_{H}$%
, where $l_{H}$ is the Hubble radius at $\eta _{0}$. Note also that the
scalar field has dimensions of (length)$^{1/2}$.

Now we consider the Klein-Gordon equation for the two phases defined above.
For simplicity, we will omit the bars, setting $\bar{k}\rightarrow k$, and $%
H_{0}\eta \rightarrow \eta $ (a dimensionless time parameter). Note that
with this parametrization, $\eta _{0}=1$ (since $H_{0}\eta _{0}=1$ in the
old variables).

\subsection{Primordial phase}

In this case, $a(\eta )=a_{0}\eta $. Hence, \emph{\ }%
\begin{equation}
\phi ^{\prime \prime }+2\frac{\phi ^{\prime }}{\eta }+k^{2}\phi =0.
\end{equation}%
The solution of this equation is 
\begin{equation}
\phi _{k}(\eta )=\eta ^{-1/2}\biggr[c_{1}\,H_{1/2}^{(1)}(k\eta
)+c_{2}\,H_{1/2}^{(2)}(k\eta )\biggl],
\end{equation}%
where $H_{\nu }^{(1,2)}(x)$ are the Hankel functions of the first and second
kind.

The Hankel functions are defined as follows: 
\begin{equation}
H_{1/2}^{(1)}(x)=J_{1/2}(x)+i\,N_{1/2}(x);H_{1/2}^{(2)}(x)=J_{1/2}(x)-i%
\,N_{1/2}(x),
\end{equation}%
where $J_{\nu }(x)$ and $N_{\nu }(x)$ are the usual Bessel and Neumann
functions. Using the fact that 
\begin{equation}
J_{1/2}(x)=\sqrt{\frac{2x}{\pi }}\frac{\sin x}{x};N_{1/2}(x)=-\sqrt{\frac{2x%
}{\pi }}\frac{\cos x}{x},
\end{equation}%
the solution for the scalar field can be re-written as 
\begin{equation}
\phi _{k}(\eta ,\vec{x})=\frac{i}{\eta }\sqrt{\frac{2}{\pi k}}\biggr[%
-c_{1}e^{i(k\eta -\vec{k}\cdot x)}+c_{2}e^{-i(k\eta +\vec{k}\cdot \vec{x})}%
\biggl].
\end{equation}%
We choose 
\begin{equation}
c_{1}=i\frac{\sqrt{\pi }}{2}\sqrt{\frac{3}{2}l_{p}},\quad c_{2}=0,
\end{equation}%
where the factor $\sqrt{\frac{3}{2}l_{pl}}$ has been chosen due to the
dimension of the scalar field for later convenience, and obtain the typical
behaviour of a normalized quantum vacuum state with a time factor added: 
\begin{equation}
\phi _{k}(\eta ,\vec{x})=\sqrt{\frac{1}{2k}}\frac{e^{i(k\eta -\vec{k}\cdot
x)}}{\eta }.
\end{equation}

\subsection{Singular phase}

In the singular phase, the Klein-Gordon equation reduces to 
\begin{equation}
\phi ^{\prime \prime }+2\phi ^{\prime }+k^{2}\phi =0.
\end{equation}%
The solutions are 
\begin{equation}
\phi =A_{+}e^{p_{+}\eta }+A_{-}e^{p_{-}\eta },
\end{equation}%
where 
\begin{equation}
p_{\pm }=-1\pm i\sqrt{k^{2}-1}.
\end{equation}%
The final solution can be written as 
\begin{equation}
\phi _{k}(\eta ,\vec{x})=e^{-\eta }\biggr[A_{+}e^{i(\omega \eta -\vec{k}%
\cdot \vec{x})}+A_{-}e^{-i(\omega \eta +\vec{k}\cdot \vec{x})}\biggl],
\label{sola}
\end{equation}%
where $\omega =\sqrt{k^{2}-1},$ and so the field propagates only if $k>H_{0}$%
. We have also incorporated the factor $\sqrt{\frac{3}{2}l_{pl}}$ in the
definition of the constants $A_{\pm }$, to make them dimensionless.

\subsection{The potential}

One way to visualizing the overall solution is to redefine the scalar field
so that 
\begin{equation}
\phi =\frac{\lambda }{a},
\end{equation}%
and then Klein-Gordon equation takes the form 
\begin{equation}  \label{potential}
\lambda ^{\prime \prime }+[k^{2}-V(a)]\lambda =0,
\end{equation}%
where $V(a)=\frac{a^{\prime \prime }}{a}$. Hence, there is propagation
whenever $k^{2}>V(a)$, and the field decays when $k^{2}<V(a)$. That is, we
have a quantum mechanical problem of particle propagation in a potential
barrier.

For the radiative universe, $a\propto \eta $, leading to $V(a)=0$ in the
primordial phase: all modes propagate \cite{grish}. In the singular phase,
however, we have a more complicated situation, since we must evaluate the
second derivative of the scale factor near the singularity. We find, 
\begin{equation}
\ddot{a}\sim -\frac{n(n-1)}{t_{s}^{2}}\biggr(\frac{y}{t_{s}}\biggl)%
^{n-2},\quad y=t_{s}-t.
\end{equation}%
Since 
\begin{equation}
\frac{\ddot{a}}{a}=\frac{a^{\prime \prime }}{a^{3}}-\frac{a^{\prime 2}}{a^{4}%
},
\end{equation}%
we find that the potential for this phase is 
\begin{equation}
V(a)=\frac{a^{\prime \prime }}{a}=-n(n-1)\frac{a_{s0}}{t_{s0}}\biggr(\frac{%
a_{s0}}{t_{s0}}\eta \biggl)^{n-2} + H_{0}^{2}.
\end{equation}%
This confirms that there is no propagation for $k<H_{0}$.

\section{Matching the solutions}

Using the solutions of the Klein-Gordon equation in the two phases, we match
the fields and their first derivative at $\eta =\eta _{0}$. This gives the
following expressions: 
\begin{eqnarray}
\sqrt{\frac{1}{2k}}e^{ik} &=&e^{-1}\biggr\{A_{+}e^{i\omega
}+A_{-}e^{-i\omega }\biggl\},  \label{match1} \\
(-1+ik)\sqrt{\frac{1}{2k}}e^{ik} &=&A_{+}(-1+i\omega )e^{(i\omega
-1)}-A_{-}(1+i\omega )e^{-(i\omega +1)}.  \label{match2}
\end{eqnarray}%
Taking the combination $((1+i\omega )\times $(\ref{match1}) + (\ref{match2}%
), and later $((1-i\omega )\times $(\ref{match1}) + (\ref{match2}), we find: 
\begin{eqnarray}
A_{+} &=&\frac{1}{2\omega }\sqrt{\frac{1}{2k}}e^{[i(k-\omega )+1]}(\omega
+k),  \label{match1bis} \\
A_{-} &=&\frac{1}{2\omega }\sqrt{\frac{1}{2k}}e^{[i(k+\omega )+1]}(\omega
-k)].  \label{match2bis}
\end{eqnarray}

\section{Energy density of created particles}

The energy density of the created particles is given by (see \cite{jerome}), 
\begin{equation}
\rho _{s}=\frac{\hbar H_{0}^{5}}{4\pi ^{2}c^{6}a^{2}}\int_{0}^{\infty
}dk\,k^{2}\biggr\{{\phi ^{\prime }}_{k}{\phi ^{\prime }}_{k}^{\ast
}+k^{2}\phi _{k}\phi _{k}^{\ast }\biggl\}.
\end{equation}%
This expression is obtained by computing the vacuum expectation value of the
Hamiltonian for a quantized massless scalar field. It is completely
equivalent to the alternative derivation of the energy density using the
Bogoliobov coefficients \cite{qrip,birr}. The pressure associated to the
created particles is given by \cite{jerome}: 
\begin{equation}
p_{s}=\frac{\hbar H_{0}^{5}}{4\pi ^{2}c^{8}a^{2}}\int_{0}^{\infty }dk\,k^{2}%
\biggr\{{\phi ^{\prime }}_{k}{\phi ^{\prime }}_{k}^{\ast }-\frac{k^{2}}{3}%
\phi _{k}\phi _{k}^{\ast }\biggl\}.
\end{equation}

Now, using (\ref{sola}) and (\ref{match1bis},\ref{match2bis}), we find the
following expression for the energy: 
\begin{eqnarray}
\rho _{s} &=&\frac{3}{2}\frac{\hbar \,H_{0}^{5}}{c^{6}}\frac{l_{pl}\,e^{y}}{%
4\pi ^{2}a_{s}^{2}}\int_{0}^{k_{m}}dk\,\frac{4k}{k^{2}-1}\biggr\{%
(2k^{2}-1)k^{2}  \notag  \label{integral} \\
&-&\cos [\sqrt{k^{2}-1}y]+\sqrt{k^{2}-1}\sin [\sqrt{k^{2}-1}y]\biggl\},
\end{eqnarray}%
where $y=2(1-\eta )$. The transition occurs at $\eta =\eta _{0}=1$ ($y=0$),
noting the re-scaling. The upper limit of integration was set by $k_{m}$, an
ultraviolet cut-off that must be identified with the Planck wavenumber. The
reason for this is that we expect the Klein-Gordon equation may not retain
its simple form (\ref{kge}) at energies higher than the Planck scale, where
quantum gravity enters. There are many studies of the modification of the
usual dispersion relation for a scalar field in the context of black hole
thermodynamics and cosmological perturbations of quantum origin, two
situations plagued by transplanckian frequencies - see \cite{jerome2} and
references therein. The modification of the dispersion relation is usually
treated by introducing a decreasing exponential term, which leads to a very
important suppression in the integration in the transplanckian region. The
adoption of this procedure is equivalent to stopping the integration near
the Planck frequency. Since the extrapolation to the transplanckian regime
is very speculative, we will ignore this transplanckian problem and adopt a
more conservative regularisation procedure \cite{birr}. We will return to
this problem later.

Let us choose the scaling so that $a_{s0}=1$. The above expressions can be
then be rewritten as 
\begin{eqnarray}  \label{energy-integral}
\rho _{s} &=&\frac{l_{pl}^{3}}{l_{H}^{3}}\frac{\rho _{0}}{\pi }%
\,e^{y}\int_{0}^{k_{m}}dk\,\frac{4k}{k^{2}-1}\biggr\{(2k^{2}-1)k^{2}  \notag
\\
&-&\cos [\sqrt{k^{2}-1}y]+\sqrt{k^{2}-1}\sin [\sqrt{k^{2}-1}y]\biggl\}\quad ,
\end{eqnarray}%
where $\rho _{0}$ is the background cosmological density at the time of the
transition.

This integral can be solved exactly and we have 
\begin{eqnarray}  \label{energy1}
\rho _{s} &=&2\frac{l_{pl}^{3}}{l_{H}^{3}}\frac{\rho _{0}}{\pi }\,e^{y}%
\biggr\{k_{m}^{4}+k_{m}^{2}+\ln [k_{m}^{2}-1]-2Ci[-\sqrt{k_{m}^{2}-1}y] 
\notag  \label{density} \\
&+&2Chi[-y]-2\frac{\cos [\sqrt{k_{m}^{2}-1}y]}{y}+2\frac{\cosh y}{y}\biggl\}.
\end{eqnarray}%
In this expression, $Ci[x]$ is the cosine integral function and $Chi[x]$ is
the hyperbolic cosine integral function.

The general behaviour indicates that the energy density of created particles
decreases, see the bottom left graphic in figure 1. This expression is
rigorously valid only after the transition. The initial condition is the
number of particles created during the first (radiative) phase. The plot is
made in units of this initial number, which is, at best, small. From this we
can conclude that the sudden singularity is robust against particle
production due to quantum effects, since the energy density of created
particles is generally much smaller than the energy density of the
background, and goes quickly to zero as the future singularity is
approached. This result also confirms the self-consistency of our
calculation using the Klein-Gordon equation to describe the evolution of a
quantum field in the metric of a classical background cosmology \cite{birr}.

If we now take the limit $k_{m}\rightarrow \infty $, keeping the
Klein-Gordon equation in its traditional form (ignoring the transplanckian
problems), we find that there is a quartic, a quadratic and a
logarithmically divergent term. Moreover, the last term in the integral (\ref%
{integral}) may only ultimately have a meaning if it is treated as a
distribution. Such divergencies may be removed using regularization of the
energy-momentum tensor. The general result depends strongly on the
background, which in our case is quite non-trivial. Even if we consider the
simplified scenario with a radiative phase preceding a singular phase, the
regularization could be implemented in the radiation-dominated phase (which
is a known result \cite{birr}), but the existence of a discontinuity in the
second derivative, due to the matching conditions we employ, makes the
application of the usual expression neither direct nor simple. But, we can
proceed in a more straightforward way. In the singular phase, we we are
evaluating the creation of particles and the scale factor becomes constant.
This makes it secure to use a covariant subtraction of the infinities \cite%
{ford}, for the density and for the pressure, 
\begin{equation}
<T_{\mu \nu }>_{reg}=<T_{\mu \nu }>_{0}-<T_{\mu \nu }>_{div}\quad ,
\end{equation}%
where $<T_{\mu \nu }>_{reg}$ stands for the regularized energy-momentum
tensor, $<T_{\mu \nu }>_{0},$ for the energy-momentum tensor evaluated using
the expressions above, and $<T_{\mu \nu }>_{div}$ is its corresponding
divergent part. The divergent parts are represented by the first term in (%
\ref{integral}). The final expression is given by (see the Appendix) 
\begin{equation}
\rho _{s}=\frac{p_{s}}{3}=4\frac{l_{pl}^{3}}{l_{H}^{3}}\frac{\rho _{0}}{\pi }%
\,\biggr\{e^{y}\,\biggr[Chi[-y]+\frac{\cosh y}{y}\biggl]\biggl\}.
\label{energy2}
\end{equation}%
It is very important to stress that the Minkowski limit, obtained by
imposing $\mathbf{\rho }_{0}\rightarrow 0$, leads to a null result, the same
as we would obtain if we had computed the vacuum expectation value in
Minkowski space-time and subtracted the divergencies. At same time, the
resulting energy-momentum is conserved. This confirms the consistency of the
procedure employed above. Note that, in the present case, the equation of
state does not coincide with the classical equation of state of a massless
scalar field in a FRW background, which is that of a stiff matter fluid ($%
p=\rho $). In general, the quantization of a classical system may change the
classical equation of state. The general form of the regularized energy
density of created particles is exhibited in figure $1$, showing a
decreasing behavior as before.

It is possible to obtain at least some information about the effects of
using a simplified model, where the evolution of the universe is described
by two phases, by integrating numerically the exact Klein-Gordon equation
using (\ref{a},\ref{b}). The results are of course plagued by the problem of
the ultraviolet divergence. But, we can stop the integration at a high
frequency and compare the result with the expression (\ref{energy1}), or
even with (\ref{energy2}). The initial condition is given by the same vacuum
state that we used above. The results for the background and for the energy
density of particles created are compared in figure $1$, where we have also
inserted the expressions for the energy of the produced particles using the
Planck frequency cut-off (\ref{energy1}) and the regularized energy-momentum
tensor (\ref{energy2}) as well as the numerical computation. Of course, the
regularized expression (\ref{energy2}) fits the numerical integration only
in a general form; this is natural since a divergent contribution has been
extracted to obtain (\ref{energy2}). Yet they agree in the sense that both
predict a decreasing energy of produced particles as the singularity is
approached. This is a quite general behavior that remains even if the free
parameters, like $a_{s0}$ and $H_{0}$, are varied. As is also expected, the
numerical fitting to the simplified analytical expression is better when the
duration of the singular phase is small compared to that of the preceding
radiative phase. Moreover, the numerical integration reveals the effect of
the sharp transition (which is not displayed in the figures): in (\ref%
{energy1},\ref{energy2}) there is a divergence in the energy at $y=0$, which
does not appear in the numerical computation. This is an effect of the
discontinuity in the second derivative, which is clearer when the
Klein-Gordon equation is written as in (\ref{potential}).

To obtain more details of the result described above, we can use the
expression for the particle production exhibited in reference \cite{birr}.
The energy density can be written as 
\begin{equation}
\rho _{s}=\frac{\hbar c}{a}\int_{0}^{\infty }N_{k}k\,d^{3}k,
\end{equation}%
where $N_{k}$ is the particle occupation number. Comparing with (\ref%
{energy-integral}), and retaining only the relevant terms after
regularization, the particle occupation is then given by 
\begin{equation}
N_{k}\propto \frac{e^{y}}{k^{2}(k^{2}-1)}\biggr\{(-\cos [\sqrt{k^{2}-1}y]+%
\sqrt{k^{2}-1}\sin [\sqrt{k^{2}-1}y]\biggl\}.
\end{equation}%
From this expression we can verify that the particle production decreases as
the sudden singularity is approached, in contrast to what happens in the
case of approach to a 'big rip' future singularity \cite{brip} (where $%
a\rightarrow \infty ,\rho \rightarrow \infty $ and $\left\vert p\right\vert
\rightarrow \infty $ as $t\rightarrow t_{s}$). Moreover, light particles are
produced more copiously than heavy particles, as would be expected. These
behaviors are shown in Figure 2.

\begin{center}
\begin{figure}[t]
\begin{minipage}[t]{0.45\linewidth}
\includegraphics[width=\linewidth]{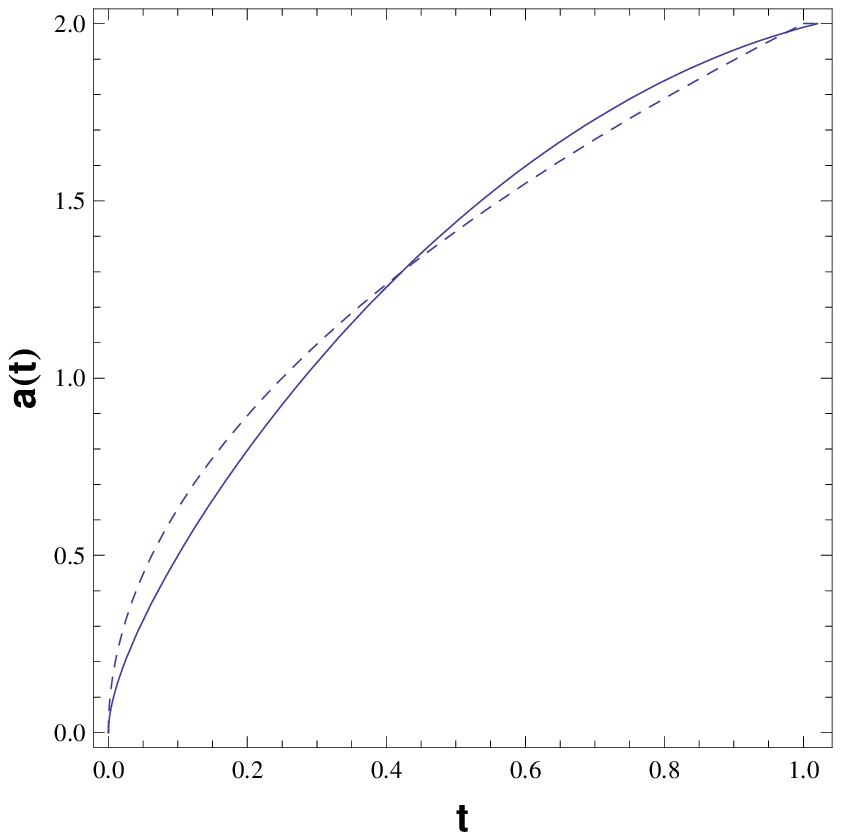}
\end{minipage} \hfill 
\begin{minipage}[t]{0.45\linewidth}
\includegraphics[width=\linewidth]{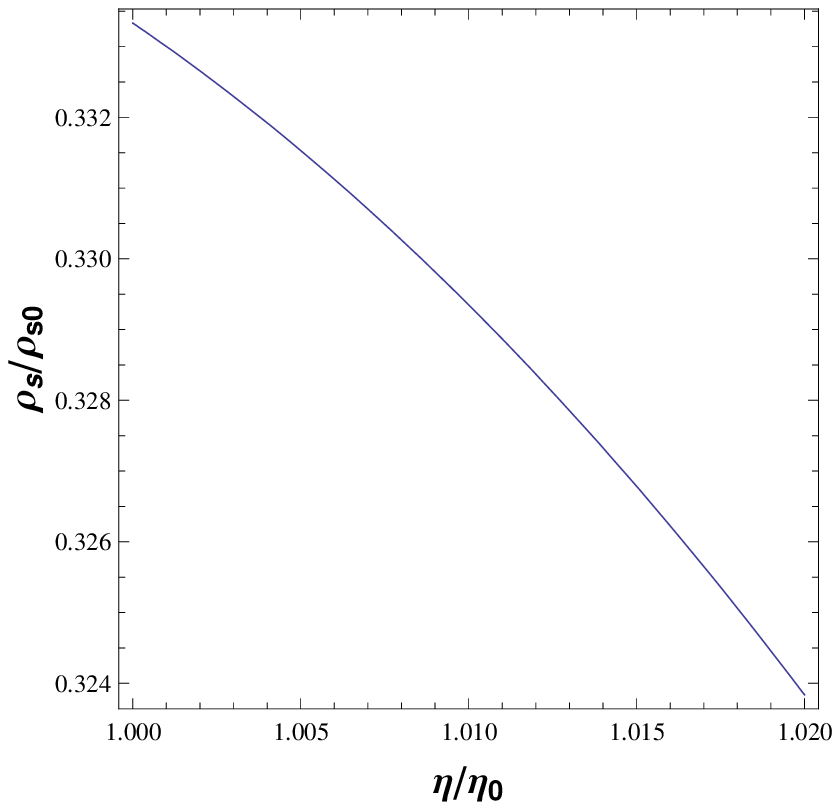}
\end{minipage} \hfill 
\begin{minipage}[t]{0.45\linewidth}
\includegraphics[width=\linewidth]{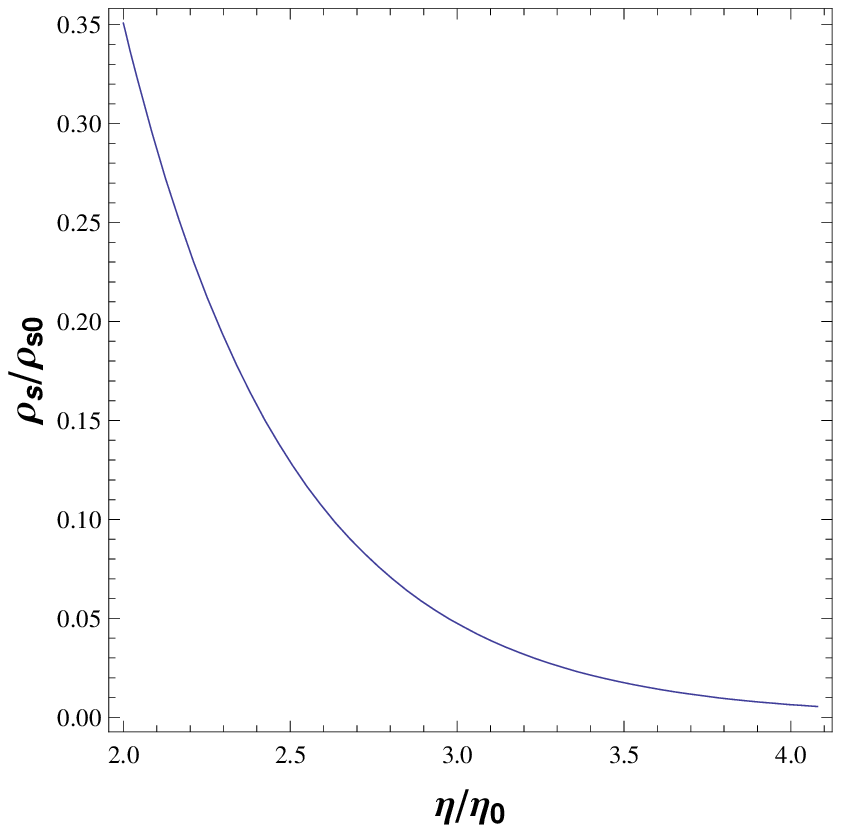}
\end{minipage} \hfill 
\begin{minipage}[t]{0.45\linewidth}
\includegraphics[width=\linewidth]{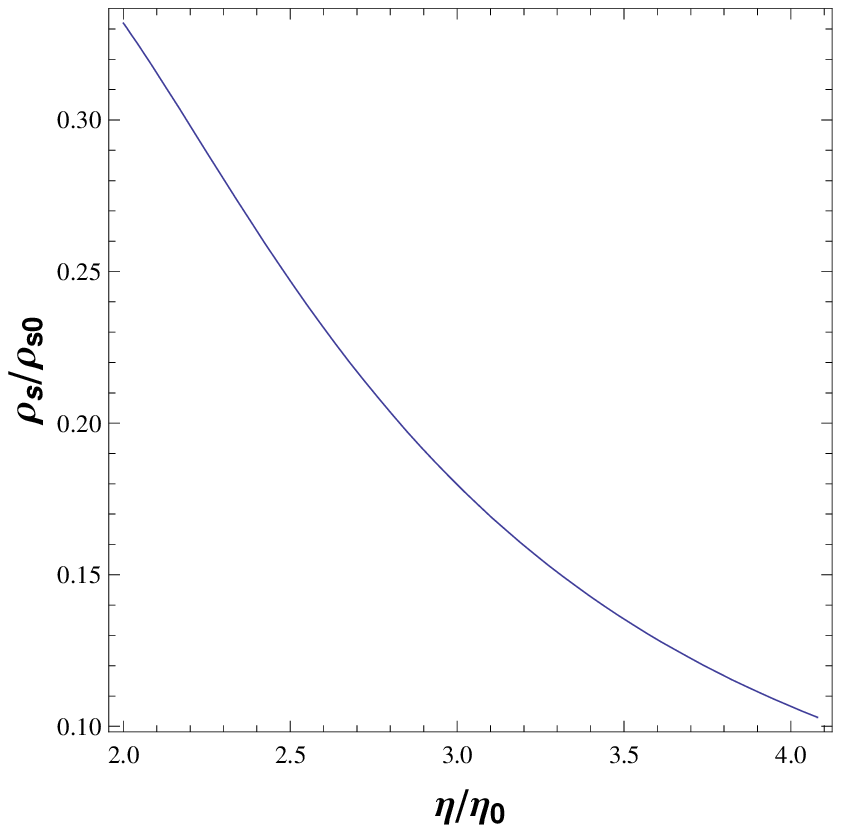}
\end{minipage} \hfill
\caption{{\protect\footnotesize These figures show the evolution of the
background scale factor (upper left) of the simplified (dashed line) and of
the exact (continuous line) models, the behavior of the energy of the
produced particles using the numerical results (upper right), the truncated
(bottom left) and the regularized (bottom right) expressions. We used
arbitrary normalization. In the figures the energies are obtained with
different time coordinates (proper time for the numerical integration and
conformal time for the analytical models) but the future direction is always
in the increasing value of the coordinate.}}
\label{}
\end{figure}

\begin{figure}[t]
\begin{minipage}[t]{0.5\linewidth}
\includegraphics[width=\linewidth]{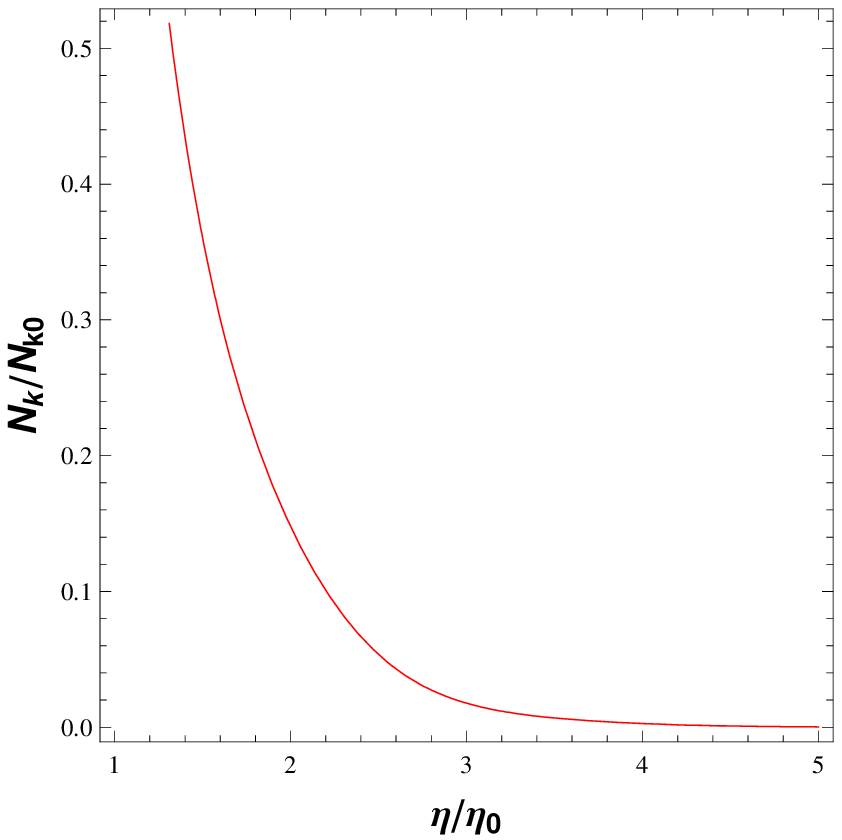}
\end{minipage} \hfill 
\begin{minipage}[t]{0.5\linewidth}
\includegraphics[width=\linewidth]{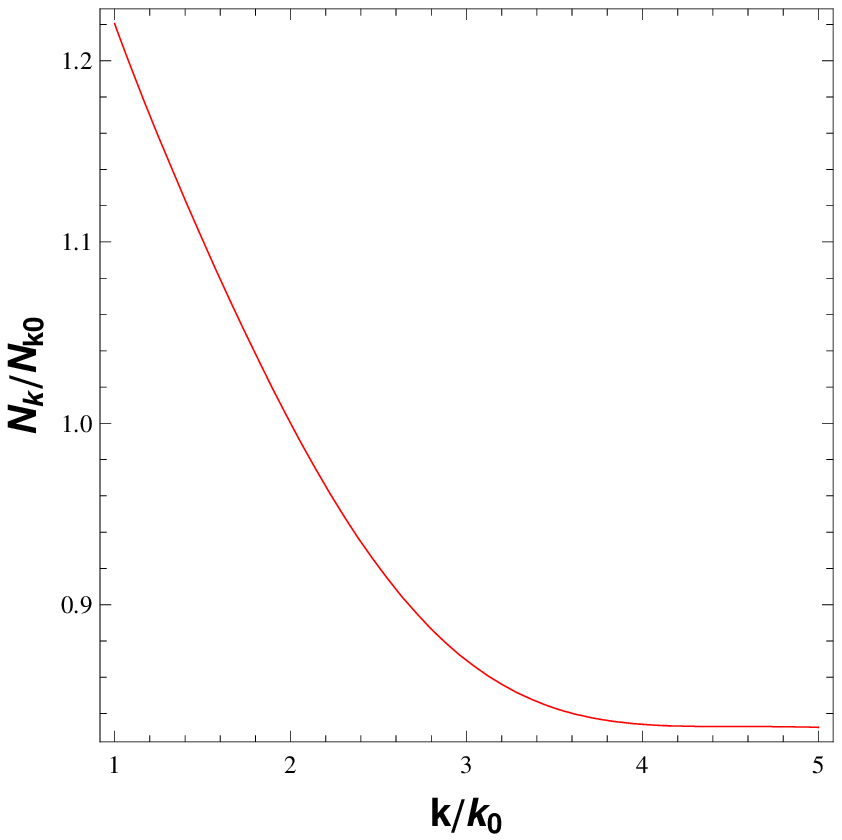}
\end{minipage} \hfill
\caption{{\protect\footnotesize The density of particles created with a
given momentum $k$ as function of time (left) and the density of particles
created as a function of the momentum $k$ for a fixed time $\protect\eta $
(right), where $k_{0}$ is a reference momentum (a value $k_{0}=2$ is chosen
arbitrarily) and $N_{k0}$ is the corresponding particle density.}}
\end{figure}
\end{center}

We can try to compare our results with those obtained in reference \cite{no}%
. However, the context is quite different to that of this paper since the
authors of \cite{no} have considered an ensemble of conformal quantum
fields, (whereas we consider only a non-conformal, massless scalar field),
generating a trace anomaly by using the effective-action approach where
gravity is modified by requiring that the quantum fields must be
renormalisable on a given metric background. They find that the sudden
singularity can be modified by quantum effects.  The energy conditions could
be violated and a big-rip singularity ensue. In our calculation, these
quantum effects are inoperative. A possible reason for these differences is
the expression used in ref. \cite{no} for the conformal anomaly: it was
derived near an initial strong curvature singularity ($a\rightarrow 0,\dot{a}%
/a\rightarrow \infty ,\rho \rightarrow \infty $ as $t\rightarrow 0$). Since
the scale factor becomes constant near the sudden singularity ($a,\dot{a}%
,\rho $ all finite but $\ddot{a}\rightarrow \infty $ as $t\rightarrow t_{s}$%
), it is possible that the expression used for the conformal anomaly does
not apply unchanged from its form in a scenario with a dynamic scale factor
near an initial singularity where the density diverges. The quantum effects
considered in reference \cite{no} have their natural domain of applicability
in the early universe, near the big bang rather than at a late-time sudden
singularity where geodesics are undisturbed. At a future sudden singularity
there is a curvature singularity which can in principle justify the
employment of a conformal anomaly for the sudden singularity. However, the
sudden singularity has many features which distinguish it from the initial
big bang singularity: the divergence appears only in the second derivative
of the scale factor (or, equivalently, in the pressure) and the expansion
rate and the density remain finite. Hence, the validity of such an
extrapolation to the sudden singularity remains to be proven. In particular,
if particle production effects from other type of fields, like massive or
conformal scalar fields, or even vectorial and spinorial quantum fields,
produce such a large back-reaction that a sudden singularity is changed into
a big rip singularity then it is difficult to describe that
self-consistently in terms of effective quantum stresses on a fixed
background because the background expansion is strongly perturbed. \newline

\subsection{Classical analogues}

The quantum particle production effects in an isotropic universe can be
viewed as arising because of the presence of an effective bulk viscosity, $%
\xi $, in the energy-momentum tensor \cite{zeld}. The presence of a
classical bulk viscosity leaves the Friedmann eq. (\ref{1}) unchanged and
modifies eq. (\ref{3}) to

\begin{equation}
\dot{\rho}+3H(\rho +p)=9H^{2}\xi ,  \label{bulk}
\end{equation}%
where $\xi (\rho )=\alpha \rho ^{m}$ on the right-hand side represents a
classical analogue of the non-equilibrium particle creation effects produced
by the viscous stress \cite{visc}. On approach to a sudden singularity, $%
a\rightarrow a_{s0},$ $H\rightarrow H_{s0}$ and $\rho \rightarrow \rho _{s0}$%
, while $\dot{\rho}$ and $\left\vert p\right\vert \rightarrow \infty $, so
we see that $\xi \rightarrow \xi _{s0}=\xi (\rho _{s0})$ and the classical
particle production term $9H^{2}\xi _{s0}$ both approach constants as $%
t\rightarrow t_{s0},$ so to leading order

\begin{equation*}
\dot{\rho}+3Hp\rightarrow 9H^{2}\xi \rightarrow \text{constant},
\end{equation*}%
and the production effects are classically negligible.

It is also interesting to ask if the presence of anisotropies in the
cosmological expansion rate could significantly change our results. In the
study of quantum effects on approach to a curvature singularity in the early
universe we are familiar with the strong effects of quantum particle
production, which can bring about significant isotropisation of the
expansion, both directly and as a result of the anisotropic red and
blue-shifting of the created particles after they become collisionless \cite%
{zeld2}. These effects are driven by the fast divergence of the dominant
shear anisotropy energy density ($\sigma ^{2}\propto a^{-6}$) at small $a$.
We could ask whether any possible amplification of small anisotropies on
approach to a sudden singularity could lead to major quantum effects.
However, we can easily see that such dominant effects will not occur on
approach to a sudden singularity as $a\rightarrow a_{s}$ when $t\rightarrow
t_{s}$. In this situation the anisotropy energy density also approaches a
constant value, $\rho (t_{s})\propto a_{s}^{-6}$ which will be far smaller
than the ambient density of dust or radiation, and the associated density of
created particles will be smaller still. There will just be small changes to
the asymptotic energy density and expansion rate in the limit $a\rightarrow
a_{s}$, as eq. (\ref{1}) is modified to

\begin{equation*}
3H_{s}^{2}=8\pi G\rho _{s0}+\sigma _{s0}^{2},
\end{equation*}%
with all three terms equal to constants. The dissipative effects of the
created particles are analogous to the presence of a shear viscous stress,
proportional to the shear and its effects also remain small as $t\rightarrow 
$ $t_{s0}$.

\section{Discussion}

We have considered the quantum particle production that would occur in an
isotropic and homogeneous universe on approach to sudden singularity, where $%
a,\dot{a}$ and $\rho $ remain finite while $\ddot{a}$ and $p$ diverge. If
significant, such quantum effects could modify or remove a sudden
singularity at finite time. We have set up a simple exactly soluble model in
which an early radiation-dominated universe evolves towards a sudden
singularity at finite time. We compute the quantum particle production as
the singularity is approached and show that its effects remain negligible
with respect to the classical background pressure and density all the way
into the singularity. We compare our results to other discussions of quantum
effects in the literature; we argue that any effects created by the presence
of small anisotropies will be negligible and show how a simple classical
description of the quantum particle production as an effective bulk
viscosity in a Friedmann universe gives a similar outcome.\vspace{0.5cm}

\noindent\textbf{Acknowledgements} A.B.B., J.C.F. and S.H. thank CNPq
(Brazil), FAPES (Brazil) and the Brazilian-French scientific cooperation
CAPES-COFECUB for partial financial support. We thank also Ilya Shapiro for
many fruitful discussions.

\section*{Appendix}

The energy density and the pressure are given by (neglecting
multiplicatively unimportant terms), 
\begin{eqnarray}
\rho _{s} &=&\frac{1}{a^{2}}e^{y}\int_{0}^{\infty }\frac{k}{\omega ^{2}}%
\biggr\{(2k-1)k^{2}-\cos (\omega y)+\omega \sin (\omega y)\biggl\}dk\quad ,
\label{pressure} \\
p_{s} &=&\frac{1}{a^{2}}e^{y}\int_{0}^{\infty }\frac{k}{\omega ^{2}}\biggr\{%
(2k-1)\frac{k^{2}}{3}-\biggr(1-\frac{2}{3}k^{2}\biggl)\cos (\omega y)+\omega
\sin (\omega y)\biggl\}dk\quad .
\end{eqnarray}%
In these expressions, $\omega =\sqrt{k^{2}-1}$ and $y=2(1-\eta )$ where $%
\eta $ is the conformal time (which is proportional to the cosmic time $t$
when $a=$ constant). Now, let us consider the conservation law, 
\begin{equation}
\rho ^{\prime }+3\frac{a^{\prime }}{a}(\rho +p)=0\quad .
\end{equation}%
The primes mean derivative with respect to $\eta $. This may be rewritten,
after redefining the time variable using the (constant) Hubble parameter, in
terms of the derivative with respect to $y$: 
\begin{equation}
-2\dot{\rho}+3(\rho +p)=0\quad .  \label{conservation}
\end{equation}%
It is easy to see that the expressions (\ref{density},\ref{pressure})
satisfy (\ref{conservation}). It is important to note that, even if $a$ is
constant, we must derive it since $\dot{a}$ (equivalently, $a^{\prime }$) is
not zero, being also equal to a constant. More importantly, if we rewrite (%
\ref{density},\ref{pressure}) as 
\begin{eqnarray}
\rho _{s} &=&\rho _{s1}+\rho _{s2}\quad , \\
p_{s} &=&p_{s1}+p_{s2}\quad ,
\end{eqnarray}%
where 
\begin{eqnarray}
\rho _{s1} &=&\frac{1}{a^{2}}e^{y}\int_{0}^{\infty }\frac{k}{\omega ^{2}}%
(2k-1)k^{2}\,dk\quad , \\
\rho _{s2} &=&\frac{1}{a^{2}}e^{y}\int_{0}^{\infty }\frac{k}{\omega ^{2}}%
\biggr\{-\cos (\omega y)+\omega \sin (\omega y)\biggl\}dk\quad , \\
p_{s1} &=&\frac{1}{a^{2}}e^{y}\int_{0}^{\infty }\frac{k}{3\omega ^{2}}%
(2k-1)k^{2}\,dk\quad , \\
p_{s2} &=&\frac{1}{a^{2}}e^{y}\int_{0}^{\infty }\frac{k}{\omega ^{2}}\biggr\{%
-\biggr(1-\frac{2}{3}k^{2}\biggl)\cos (\omega y)+\omega \sin (\omega y)%
\biggl\}dk\quad ,
\end{eqnarray}%
then the pairs ($\rho _{s1},p_{s2}$) and ($\rho _{s1},p_{s2}$) satisfy (\ref%
{conservation}) separately.

Now we can easily show that $\rho _{s2}$ is finite. In fact,it can be
written as 
\begin{equation}
\rho _{s2}=-\frac{1}{a^{2}}\frac{d}{dy}\int_{0}^{\infty }\frac{k\,dk}{\omega
^{2}}\biggr[e^{y}\cos (\omega y)\biggl]\quad .
\end{equation}%
The integral can now be performed, leading to 
\begin{equation}
\rho_{s2}=\frac{1}{a^{2}}\frac{d}{dy}\biggr(e^{y}Chi(-y)\biggl)\quad ,
\end{equation}%
where we have exploited the fact that the cosine function is even in its
argument. This function is finite, except at $y=1$, an effect of the sharp
transition of the second derivative in the simplified model which does not
appear in the numerical integration, where the background is smooth.

Now we can obtain the pressure $p_{s2}$. The result is 
\begin{equation}
p_{s2}=\frac{1}{3}\frac{e^{y}}{a^{2}}\biggr\{Chi(-y)+3\frac{\cosh y}{y}-2%
\frac{\cosh y}{y^{2}}+2\frac{\sinh y}{y}\biggl\}\quad ,
\end{equation}%
which goes to zero asymptotically.

Hence, after subtracting the divergent parts $\rho _{s1}$ and $p_{s1}$, we
obtain finite expressions for the energy and for the pressure, which have
the ordinary Minkowski space-time limit satisfying the covariant
energy-momentum conservation law.

The regularization procedure is employed here is inspired in the $n$-wave
technique \cite{zeld} (see also \cite{pavlov}). Here, its application is
very simple due to the natural splitting of the energy density and pressure
into a divergent and a finite parts, which separately satisfy the
conservation law.

\end{document}